\documentclass[11pt]{article}

\usepackage{latexsym}
\usepackage{amsfonts}

\newtheorem{theorem}{Theorem}

\begin{document}

\title{The Einstein-Vlasov system/Kinetic theory} 
\author{H\aa kan Andr\'{e}asson\\
        Department of Mathematics\\
        Chalmers University of Technology\\
        S-412\,96 G\"oteborg, Sweden\\
        hand@math.chalmers.se\\
        http://www.math.chalmers.se/\~{}hand} 

\date{}
\maketitle

\begin{abstract}
The main purpose of this article is to provide a guide to theorems on 
global properties of solutions to the Einstein-Vlasov system. This
system couples Einstein's equations to a kinetic matter model. Kinetic 
theory has been an important field of research during several decades
in which the main focus has been on nonrelativistic and special
relativistic physics, {\it i.e.} to 
model the dynamics of neutral gases, plasmas, and Newtonian
self-gravitating systems. In 1990, Rendall and Rein initiated a
mathematical study of the Einstein-Vlasov system. 
Since then many theorems on global properties of solutions to this 
system have been established. The Vlasov equation describes matter 
phenomenologically and it should be stressed that most of the theorems
presented in this article are not presently known for other such
matter models ({\it i.e.} fluid models). This paper gives
introductions to kinetic theory in non-curved spacetimes and then the
Einstein-Vlasov 
system is introduced. We believe that a good understanding of kinetic
theory in non-curved spacetimes is fundamental to good 
comprehension of kinetic theory in general relativity. 
\end{abstract}

\newpage
\section{Introduction to kinetic theory} 
In general relativity, kinetic theory has been used
relatively sparsely to model phenomenological matter in comparison to 
fluid models. From a mathematical point of view, however, there are
fundamental advantages to using a kinetic description. In 
non-curved spacetimes kinetic theory has been studied intensively as a
mathematical subject during several decades and it has also played an
important role from an engineering point of view. In the first part of
this introduction, we will review kinetic theory in non-curved 
spacetimes and we will consider mainly the special relativistic case,
but mathematical results in the nonrelativistic case
will also be discussed. We believe that 
a good understanding of kinetic theory in non-curved spacetimes is
fundamental to good comprehension of kinetic theory in general relativity. 
Moreover, it is often the case that mathematical methods used to treat
the Einstein-Vlasov system are carried over from methods developed in
the special relativistic or nonrelativistic case. 

The purpose of kinetic theory is to model the time evolution of a
collection of particles. The particles may be entirely different 
objects depending on the physical situation. For instance, 
the particles are 
atoms and molecules in a neutral gas or electrons and ions in a 
plasma. In stellar dynamics the particles are stars and in a 
cosmological case they are galaxies or even clusters 
of galaxies. 
Mathematical models of particle systems are most frequently described 
by kinetic or fluid equations. 
A characteristic feature of kinetic theory is 
that its 
models are statistical and the particle systems are described 
by distribution functions $f=f(t,x,p)$, which 
represent the density of particles with given space-time position 
$(t,x)\in\mathbb{R}\times\mathbb{R}^3$ and 
momentum $p\in \mathbb{R}^{3}$. A distribution function 
contains a wealth of information, and macroscopical 
quantities are easily calculated from this function. 
In a fluid model the quantities that describe the
system do not depend on the momentum $p$ but only on 
the spacetime 
point $(t,x)$. A choice of
model is usually made with regard to the physical properties of
interest for the system or with regard to numerical considerations. 
It should be mentioned that a fluid model that is too naive may give rise to
shell-crossing singularities, which are unphysical. 
In a kinetic description such phenomena are ruled out. 

The time evolution of the system is determined by the interactions 
between the particles which depend on the physical situation. For
instance, the driving mechanism for the time evolution of a neutral
gas is the collision between particles (the relativistic
Boltzmann equation). For a plasma the interaction is 
through the electric charges (the Vlasov-Maxwell system) 
and in the stellar and 
cosmological cases the interaction is gravitational (the
Einstein-Vlasov system). Of course, combinations of interaction
processes are also considered but in many situations one of them is
strongly dominating and the weaker processes are neglected. 

\subsection{The relativistic Boltzmann equation}
Consider a collection of neutral particles in Minkowski spacetime. Let 
the signature of the metric be $(-,+,+,+)$, let all the particles 
have rest mass $m=1$, and normalize the speed of light, $c$, to one. 
The four-momentum of a particle is denoted by 
$p^{a}$, $a=0,1,2,3$. 
Since all particles have equal rest mass, the four-momentum for
each particle 
is restricted to the mass shell, 
$p^{a}p_{a}=-m^{2}=-1$. Thus, by denoting the three-momentum 
by $p\in\mathbb{R}^{3}$, $p^{a}$ may be written 
$p^{a}=(-p^0,p)$, where $|p|$ is the usual 
Euclidean length of $p$ and $p^{0}=\sqrt{1+|p|^{2}}$ is the energy of
a particle with three-momentum $p$. The relativistic 
velocity of a particle with momentum $p$ is denoted by $\hat{p}$ and
is given by 
\begin{equation}
\hat{p}=\frac{p}{\sqrt{1+|p|^{2}}}.\label{velo}
\end{equation}
Note that $|\hat{p}|<1=c$. 
The relativistic Boltzmann equation models the space-time behaviour 
of the one-particle distribution function $f=f(t,x,p),$ and it 
has the form 
\begin{equation}
(\partial _{t} + \frac{p}{p^{0}}\cdot\nabla _{x})f=Q(f,f), \label{rbe}
\end{equation}
where the relativistic collision operator $Q(f,g)$ is defined by 
\begin{eqnarray}
&\displaystyle Q(f,g)=\int_{\mathbb{R}^{3}}\int_{\mathbb{S}^{2}} k(p,q,\omega)\times
&\nonumber\\ &\displaystyle 
\times [f(p+a(p,q,\omega)\omega) g(q-a(p,q,\omega)\omega)-f(p)g(q)]
dpd\omega. &\label{rbeQ} 
\end{eqnarray}
(Note that $g=f$ in (\ref{rbe})). 
Here $d\omega$ is the element of surface area on $\mathbb{S}^{2}$ and 
$k(p,q,\omega)$ is the scattering kernel, which depends on 
the scattering cross-section in the interaction process. See~\cite{GLW} for
a discussion about the scattering kernel. The function 
$a(p,q,\omega)$ results from the collision mechanics. 
If two particles, with momentum $p$ and $q$ respectively, collide 
elastically (no energy loss) with scattering 
angle $\omega\in\mathbb{S}^{2}$, their momenta will 
change, $p\rightarrow p'$ and $q\rightarrow q'$. 
The relation between $p,q$ and $p',q'$ is 
\begin{equation}
p'=p+a(p,q,\omega)\omega,\,\,\,q'=q-a(p,q,\omega)\omega,
\end{equation}
where 
\begin{equation}
a(p,q,\omega)=\frac{2(p^{0}+q^{0})p^{0}
q^{0}(\omega\cdot(\hat{q}-\hat{p}))}
{(p^{0}+q^{0})^{2}-(\omega\cdot(p+q))^{2}}.
\end{equation}
This relation is a consequence of four-momentum conservation, 
$$p^{a}+q^{a}=p^{a'}+q^{a'},$$ or equivalently 
\begin{eqnarray}
p^{0}+q^{0}&=&p^{0'}+q^{0'},\\ 
p+q&=&p'+q'.
\end{eqnarray}
These are the conservation equations for relativistic particle
dynamics. In the classical case 
these equations read 
\begin{eqnarray}
|p|^{2}+|q|^{2}&=&|p'|^{2}+|q'|^{2},\label{cocl1}\\
p+q&=&p'+q'\label{cocl2}. 
\end{eqnarray}
The function $a(p,q,\omega)$ is the distance between 
$p$ and $p'$ ($q$ and $q'$) and the analogue function 
in the Newtonian case has the form 
\begin{equation}
a_{cl}(p,q,\omega)=\omega\cdot (q-p). 
\end{equation}
By inserting $a_{cl}$ in place of $a$ in (\ref{rbeQ}) we obtain 
the classical Boltzmann collision operator (disregarding 
the scattering kernel, which is also different). In~\cite{C1}
classical solutions to the relativistic Boltzmann equations are
studied as $c\to\infty,$ and it is proved that the limit as
$cto\infty$ of these solutions satisfy the classical Boltzmann equation. 

The main result concerning the existence of solutions to the classical
Boltzmann equation is a theorem by DiPerna and Lions~\cite{DL1} that proves 
existence, but not uniqueness, of renormalized solutions 
(i.e. solutions in a weak sense, which are even more 
general than distributional solutions). An analogous result holds in
the relativistic case, as was shown by Dudy\'{n}sky and
Ekiel-Jezewska~\cite{DE}. Regarding classical solutions, 
Illner and Shinbrot~\cite{IlS} have shown global existence of solutions to
the nonrelativistic Boltzmann equation for small initial data (close to
vacuum). At present there is no analogous result for the relativistic 
Boltzmann equation and this must be regarded as an interesting open
problem. There is however a recent result~\cite{NT} for the homogeneous
relativistic Boltzmann equation where global existence is shown for
small initial data and bounded scattering kernel. 
When the data are 
close to equilibrium (see below), global existence of classical
solutions has been proved by Glassey and Strauss~\cite{GSt3} in the relativistic 
case and by Ukai~\cite{U} in the nonrelativistic case (see
also~\cite{Sh} and~\cite{NI}). 

The collision operator $Q(f,g)$ may be written in an obvious way as 
$$Q(f,g)=Q^{+}(f,g)-Q^{-}(f,g),$$ where $Q^{+}$ and $Q^{-}$ are
called the gain and loss term respectively. If the loss term is
deleted the gain-term-only Boltzmann equation is obtained. It is
interesting to note that the methods of proof for the small data
results mentioned above concentrate on gain-term only equations and
once that is solved it is easy to include the loss term. In~\cite{ACI}
it is shown that the gain-term only classical and relativistic
Boltzmann equations blow up for initial data not restricted to a small
neihgbourhood of trivial data. Thus, if a global existence proof for
unrestricted data will be given it will necessarily use the full
collision operator. 

The gain term has a nice regularizing property in the momentum variable. 
In~\cite{An1} it is proved that given 
$f\in L^{2}(\mathbb{R}^{3})$ and $g\in L^{1}(\mathbb{R}^{3})$ with
$f,g\geq 0$, then  
\begin{equation}
\| Q^{+}(f,g)\|_{H^{1}(\mathbb{R}_{p}^{3})}\leq C\| f\|_{L^{2}
(\mathbb{R}_{p}^{3})}\|g\|_{L^{1}(\mathbb{R}_{p}^{3})},\label{Q} 
\end{equation}
under some technical requirements on the scattering kernel. 
Here $H^{s}$ is the usual Sobolev space. 
This regularizing result was first proved by 
P.L. Lions~\cite{Li} in the classical situation. 
The proof relies on the theory of Fourier integral
operators and on the method of stationary phase, and requires 
a careful analysis of the collision geometry, which is very different 
in the relativistic case. See also~\cite{Wen1} and~\cite{Wen2} for a 
simplified proof in the classical and relativistic case respectively. 

The regularizing theorem has many applications. An important
application is to prove that solutions tend to equilibrium for large
times. More precisely, Lions used the regularizing theorem
to prove that solutions to the (classical) Boltzmann 
equation, with periodic boundary conditions, converge in $L^{1}$ to 
a global Maxwellian, $$M=e^{-\alpha |p|^{2}+\beta\cdot p+\gamma}\;\;
\alpha,\gamma\in R,\;\alpha>0,\;\beta\in \mathbb{R}^{3},$$ 
as time goes to infinity. This result had first been obtained by
Arkeryd~\cite{Ar} by using non-standard analysis. It should be pointed out
that the convergence takes place through a sequence of times tending
to infinity and it is not known whether the limit is unique or depends on
the sequence. 
In the relativistic situation, the analogous question of 
convergence to
a relativistic Maxwellian, or a J\"{u}ttner 
equilibrium solution, 
$$J=e^{-\alpha\sqrt{1+|p|^{2}}+\beta\cdot
  p+\gamma},\;\;\alpha,\beta\mbox{ and }\gamma\mbox
  { as above, with } \alpha>|\beta|,$$ 
had been studied by Glassey and Strauss~\cite{GSt3},~\cite{GSt4}. 
In the periodic
case they proved 
convergence in a variety of function spaces for initial data close to
a J\"{u}ttner solution. 
Having obtained the regularizing theorem for the relativistic 
gain term, it is a straightforward task to follow the method of Lions
and prove convergence to 
a \textit{local} J\"{u}ttner solution for 
arbitrary data (satisfying the natural
bounds of finite energy and entropy) that is periodic in the space
variables. In~\cite{An1} it is next proved that 
the local J\"{u}ttner solution must be a global one, due to the
periodicity of the solution. 

For more information on the relativistic Boltzmann equation on 
Minkowski space we refer to~\cite{Gl},~\cite{GLW}, and ~\cite{Sy} and in the nonrelativistic
case we refer to the excellent review paper by Villani~\cite{V} and the
books~\cite{Gl},~\cite{CIP}. 
\subsection{The Vlasov-Maxwell- and Vlasov-Poisson systems} 
Let us consider a collisionless plasma, which is a collection of 
particles for which  collisions
are relatively rare and the interaction is through 
their charges. We assume below that the plasma consists only of one type of
particle, whereas the results below also hold for plasmas with several
particle species. 
The particle rest mass is normalized to one. 
In the kinetic framework, the most general set of equations for modelling a
collisionless plasma is the relativistic
Vlasov-Maxwell system: 
\begin{equation}
\partial_{t}f+\hat{v}\cdot\nabla_{x}f+(E(t,x)+\hat{v}\times
B(t,x))\cdot\nabla_{v}f=0\label{rvm}
\end{equation}
\begin{eqnarray}
  \label{maxwell}
  &\partial_{t}E+j=c\nabla\times B,\;\;\;\;\;\nabla\cdot E=\rho, &\\ 
  &\partial_{t}B=-c\nabla\times E,\;\;\;\;\;\nabla\cdot B=0.&\label{maxwell_2}
\end{eqnarray}
The notation follows the one already introduced with the exception
that the momenta are now denoted by $v$ instead of $p$. This has become
a standard notation in this field. 
$E$ and $B$ are the
electric and magnetic fields, and $\hat{v}$ is the
relativistic velocity 
\begin{equation}
\hat{v}=\frac{v}{\sqrt{1+|v|^2/c^2}},
\label{relveldef}
\end{equation}
where $c$ is the speed
of light. 
The charge density $\rho$ and current $j$ are given by 
\begin{equation}
  \rho=\int_{\mathbb{R}^{3}} fdv,\;\;\;\;\; j=\int_{\mathbb{R}^{3}}
\hat{v}fdv. 
\end{equation}
Equation (\ref{rvm}) is the relativistic Vlasov equation and 
(\ref{maxwell}) and (\ref{maxwell_2}) are the Maxwell equations. 

A special case in three dimensions is obtained by considering 
spherically symmetric initial data. For such data it can be shown 
that the solution also will be spherically symmetric, and that 
the magnetic field has to be constant. The Maxwell equation 
$\nabla\times E=-\partial_tB$ then implies that the electric field 
is the gradient of a potential $\phi$. 
Hence, in the spherically
symmetric case the relativistic Vlasov-Maxwell system 
takes the form:
\begin{eqnarray}
&\partial_{t}f+\hat{v}\cdot\nabla_{x}f+\beta E(t,x)
\cdot\nabla_{v}f=0&\label{rv}\\
&E=\nabla\phi,\;\;\;\Delta\phi=\rho.&\label{rpoisson}
\end{eqnarray}
Here $\beta=1,$ and the constant magnetic field has been 
set to zero, since a constant field has no significance in 
this discussion. 
This system makes sense
for any initial data, without symmetry constraints, and 
is called the relativistic Vlasov-Poisson equation. 
Another 
special case of interest is the classical limit, 
obtained by letting $c\rightarrow\infty$ in
(\ref{rvm})-(\ref{maxwell_2}), yielding: 
\begin{eqnarray}
&\partial_{t}f+v\cdot\nabla_{x}f+\beta E(t,x)
\cdot\nabla_{v}f=0&\label{v}\\
&E=\nabla\phi,\;\;\;\Delta\phi=\rho,&\label{poisson}
\end{eqnarray}
where $\beta=1.$ 
See Schaeffer~\cite{Sc1} for a rigorous derivation of this result. 
This is the (nonrelativistic) Vlasov-Poisson equation, 
and $\beta=1$ corresponds to repulsive 
forces (the plasma case). Taking $\beta=-1$ means attractive forces
and the Vlasov-Poisson equation is then a model for a Newtonian
self-gravitating system. 

One of the fundamental problems in kinetic theory is to find out
whether or not spontaneous shock formations will develop in a
collisionless gas, i.e. whether solutions 
to any of the equations above will remain smooth 
for all 
time, given smooth initial data. 

If the initial data are small this problem has an affirmative solution
in all 
cases considered above (see~\cite{GSc1},~\cite{GSt2},~\cite{BD}, and~\cite{BDH}). 
For initial data unrestricted in size the picture is more involved. 
In order to 
obtain smooth solutions globally in time, the main issue is to 
control the support of the momenta 
\begin{equation}
Q(t):=\sup \{|v|:\exists (s,x)\in [0,t]\times \mathbb{R}^3\mbox{ such that
    }f(s,x,v)\not= 0\},\label{Q}
\end{equation} 
i.e. to bound $Q(t)$ by a continuous function so that $Q(t)$ will not
blow up in finite time. 
That such a control is sufficient for obtaining global existence of
smooth solutions follows from well-known results in the different
cases (see~\cite{GSt1},~\cite{Ho1},~\cite{B} and~\cite{GSc1}). 
For the full three-dimensional relativistic Vlasov-Maxwell system, this
important problem of establishing whether or not solutions will remain
smooth for all time is open. There has been an increased activity
during the last years aiming at a solution of this problem. Two new 
methods of proofs of the same theorem as in~\cite{GSt1} are 
given in~\cite{KS} and~\cite{BGP}. A different sufficient criteria for
global existence is given by Pallard in~\cite{P1} and he also shows a 
new bound for the electromagnetic field in terms of $Q(t)$
in~\cite{P2}. 
In two space and three momentum 
dimensions, Glassey and Schaeffer~\cite{GSc2} have shown that $Q(t)$ can be
controlled, which thus yields global existence of smooth solutions 
in that case (see also~\cite{GSc3}). 

The relativistic and nonrelativistic Vlasov-Poisson equations 
are very similar in form. 
In particular, the equation for the field is 
identical in the two cases. 
However, the mathematical results concerning the two systems are 
very different. 
In the nonrelativistic case Batt~\cite{B} gave an affirmative solution 1977
in the case of spherically symmetric data. Pfaffelmoser~\cite{Pf} (see
also Schaeffer~\cite{Sc4}) was the first one to give a proof for general
smooth data. He obtained the bound 
$$Q(t)\leq C(1+t)^{(51+\delta)/11},$$ where $\delta>0$ could be taken
arbitrarily small. This bound was later improved by different authors. 
The sharpest bound valid 
for $\beta=1$ and $\beta=-1$ has been given by Horst~\cite{Ho2} and reads 
$$Q(t)\leq C(1+t)\log(2+t).$$ In the case of repulsive forces 
($\beta=1$) Rein~\cite{Rn1} has found the sharpest estimate by using a new
identity for the Vlasov-Poisson equation, discovered independently 
by Illner and Rein~\cite{IlR} and by Perthame~\cite{Pe}. Rein's estimate reads 
$$Q(t)\leq C(1+t)^{2/3}.$$ Independently and about the same time as 
Pfaffelmoser gave his proof, Lions and Perthame~\cite{LP} used a different
method for proving global existence. To some extent their method
seems to be more generally applicable to attack problems similar to
the Vlasov-Poisson equation but which are still quite different 
(see~\cite{An2},~\cite{KR}). On the other hand, their method does not give such
strong growth estimates on $Q(t)$ as described above. 
For the relativistic Vlasov-Poisson equation, 
Glassey and Schaeffer~\cite{GSc1} showed in the case $\beta=1$ that if
the data are spherically symmetric, $Q(t)$ can be controlled, which is
analogous to 
the result by Batt mentioned above. Also in the case of cylindrical
symmetry they are able to control $Q(t),$ see~\cite{GSc4}. If $\beta=-1$ it was
also shown in~\cite{GSc1} 
that blow-up occurs in finite time for spherically symmetric
data with negative total energy. This system, however, is unphysical in the
sense that it is not a
special case of the Einstein-Vlasov system. 
Quite surprisingly, for general smooth initial data 
none of the techniques discussed above for the
nonrelativistic Vlasov-Poisson equation apply in the relativistic case. 
This fact is annoying since it has 
been suggested 
that an understanding of this equation may be necessary 
for understanding the three-dimensional relativistic Vlasov-Maxwell 
equation. However, the relativistic Vlasov-Poisson equation 
lacks the Lorentz invariance;
it is a hybrid of a classical Galilei invariant field equation and a
relativistic transport equation (\ref{rv}). Only for spherical
symmetric data is the equation a fundamental 
physical equation. The classical Vlasov-Poisson equation 
is on the other hand Galilean invariant. 
In~\cite{An3} a different equation for the field is introduced that is observer 
independent
among Lorentz observers. 
By coupling this equation for the field 
to the relativistic Vlasov equation, the function $Q(t)$ 
may be controlled as shown in~\cite{An3}. 
This is an indication that the transformation
properties are important in studying existence of smooth
solutions (the situation is less subtle for weak solutions, 
where 
energy estimates and averaging are the main tools, see~\cite{HH},
~\cite{DL2} and~\cite{Rn7}). 
Hence, it
is unclear whether or not the relativistic Vlasov-Poisson equation will 
play a central role 
in the understanding of the Lorentz invariant relativistic 
Vlasov-Maxwell equation. 

We refer to the book by Glassey ~\cite{Gl} for more information on the
relativistic Vlasov-Maxwell system and the Vlasov-Poisson equation. 
\subsection{The Nordstr\"{o}m-Vlasov system} 
Before turning to the main theme of this review, i.e. the Einstein-Vlasov
system, we introduce a system which has many mathematical features in
common with the Vlasov-Maxwell system and which recently has been
mathematically studied for the first time. It is based on 
an alternative theory of gravity which was given by Nordstr\"{o}m~\cite{Nor}
in 1913. By coupling this model to a kinetic description of matter
the Nordstr\"{o}m-Vlasov system results. In Nordstr\"{o}m gravity
the scalar field $\phi$ contains the gravitational effects as 
described below. The Nordstr\"{o}m-Vlasov system reads 
\begin{equation} \label{wave1}
\partial_t^2\phi-\bigtriangleup_x\phi=-e^{4\phi}
\int\mathfrak{f}\,\frac{dp}{\sqrt{1+|p|^{2}}},
\end{equation}
\begin{equation} \label{vlasov1}
\partial_{t}\mathfrak{f} + \widehat{p}\cdot\nabla_x\mathfrak{f} -
\left[\left(\partial_{t}\phi + \widehat{p}\cdot\nabla_x\phi \right) p +
(1+|p|^{2})^{-1/2}\nabla_x\phi\right]\cdot\nabla_p\mathfrak{f}=0.
\end{equation}
Here  
\[
\widehat{p} =\frac{p}{\sqrt{1+|p|^2}}
\]
denotes the relativistic velocity of a particle with momentum $p$. 
The mass of each particle, 
the gravitational constant and the speed of light are 
all normalized to one. 
A solution $(\mathfrak{f},\phi)$ of this system 
is interpreted as follows: The spacetime is a 
Lorentzian manifold with a conformally flat metric which, in the coordinates 
$(t,x)$, takes the form 
\[
g_{\mu\nu}=e^{2\phi} \textrm{diag}(-1,1,1,1).
\]
The particle distribution $f$ defined on the mass shell 
in this metric is given by
\begin{equation} \label{fph}
f(t,x,p)=\mathfrak{f}(t,x,e^\phi p).
\end{equation}
The first mathematical study of this system was undertaken by
Calogero in~\cite{C2} 
where static solutions are analyzed and where also more details on the
derivation of the system are given. Although the Nordstr\"{o}m-Vlasov 
model of gravity does not describe physics correctly (in the classical
limit the Vlasov-Poisson system of Newtonian gravity~\cite{CL} is
however obtained) it can nevertheless serve as a 
platform for developing new mathematical methods. 
The system has some common 
features with the Einstein-Vlasov system (see the next section) but
seems in many respects to be less involved. The closest 
relationship from a mathematical point of view is the Vlasov-Maxwell
system; this is evident in~\cite{CR1} 
where weak solutions are obtained in analogy with~\cite{DL2}, 
in~\cite{CR2} where a continuation criterion for existence of classical
solutions is established in analogy with~\cite{GSt1} (an alternative
approach is given in~\cite{P3}) and in~\cite{Le3} where global 
existence in the case of two space dimensions is proved 
in analogy with~\cite{GSc3}. 
The spherically 
symmetric case is studied in~\cite{ACR} and cannot be directly
compared to the spherically symmetric Vlasov-Maxwell system (i.e. the
Vlasov-Poisson system) since the hyperbolic nature of the equations is
still present in the former system while it is lost in the latter
case. In~\cite{ACR} it is shown that
classical solutions exist globally in time for compactly supported
initial data under the additional condition that there is a lower
bound on the modulus of the angular momentum of the initial particle
system. It should be noted that this is not a smallness condition and
that the result holds for arbitrary large initial data satisfying this 
hypothesis. 
\subsection{The Einstein-Vlasov system} 
In this section we will consider a self-gravitating collisionless gas
and we will write down the Einstein-Vlasov system and describe its
general mathematical features. Our 
presentation follows to a large extent the one by Rendall in~\cite{Rl6}. We 
also refer to 
Ehlers~\cite{E} and Stewart~\cite{St} for more background on kinetic theory in general
relativity. 

Let $M$ be a four-dimensional manifold and let $g_{ab}$ be a metric
with Lorentz signature $(-,+,+,+)$ so that $(M,g_{ab})$ is a
spacetime. We use the abstract index notation, which means that
$g_{ab}$ is a geometric object and not the components of a tensor. 
See~\cite{Wa} for a discussion on this notation. The metric is assumed to 
be time-orientable so that there is a distinction between future and past
directed vectors. The worldline of a particle with non-zero rest mass $m$ 
is a timelike curve and the unit future-directed tangent vector $v^a$
to this curve is the four-velocity of the particle. The four-momentum $p^a$ is
given by $mv^a.$  We assume that all particles have equal rest mass
$m$ and we normalize so that $m=1$. One can also consider massless
particles but we will rarely discuss this case. The possible values of
the four-momentum are all future-directed unit timelike vectors and
they constitute a hypersurface $P$ in the tangent bundle $TM$, which is 
called the mass shell. The distribution function $f$ that we introduced in
the previous sections is a non-negative function on $P.$ Since we are
considering a collisionless gas, the particles travel along geodesics
in spacetime. The Vlasov equation is an equation for $f$ that exactly 
expresses this fact. To get an explicit expression for this equation
we introduce local coordinates on the mass shell. We choose local
coordinates on $M$ such that the hypersurfaces $t=x^0=$ constant are 
spacelike so that $t$ is a time coordinate and $x^j,\;j=1,2,3$ are
spatial coordinates (letters in the beginning of the alphabet always
take values $0,1,2,3$ and letters in the middle take $1,2,3$). 
A timelike vector is future directed if and only if its zero component
is positive. Local coordinates on $P$ can then 
be taken as $x^a$ together with the spatial components of the 
four-momentum $p^a$ in these coordinates. 
The Vlasov equation then reads 
\begin{equation}
\partial_{t}f+\frac{p^{j}}{p^0}\partial_{x^j}f
-\frac{1}{p^0}\Gamma^{j}_{ab}p^ap^b\partial_{p^j}f=0.\label{vlasovgamma} 
\end{equation} 
Here $a,b=0,1,2,3$ and $j=1,2,3$ and $\Gamma^{j}_{ab}$ are the
Christoffel symbols. It is understood that $p^0$ is expressed in terms
of $p^j$ and the metric $g_{ab}$ using the relation $g_{ab}p^ap^b=-1$
(recall that $m=1$). 

In a fixed spacetime the Vlasov equation is a linear hyperbolic
equation for $f$ and we can solve it by solving the characteristic
system, 
\begin{eqnarray} 
dX^i/ds&=&\frac{P^i}{P^0},\label{char1}\\ 
dP^i/ds&=&-\Gamma^i_{ab}\frac{P^aP^b}{P^0}.\label{char2} 
\end{eqnarray} 
In terms of initial data $f_0$ the solution to the Vlasov equation can
be written 
\begin{equation}
f(x^a,p^i)=f_0(X^i(0,x^a,p^i),P^i(0,x^a,p^i)),\label{solution} 
\end{equation} 
where $X^i(s,x^a,p^i)$ and $P^i(s,x^a,p^i)$ solve (\ref{char1}) and
(\ref{char2}) and where $X^i(t,x^a,p^i)=x^i$ and $P^i(t,x^a,p^i)=p^i.$ 

In order to write down the Einstein-Vlasov system we need to define
the energy-momentum tensor $T_{ab}$ in terms of $f$ and $g_{ab}.$ 
In the coordinates $(x^a,p^a)$ on $P$ we define 
\begin{displaymath} 
T_{ab}=-\int_{\mathbb{R}^{3}}f\,p_ap_b|g|^{1/2}\frac{dp^1dp^2dp^3}{p_0}, 
\end{displaymath} 
where as usual $p_a=g_{ab}p^b$ and $|g|$ denotes the absolute value of
the determinant of $g.$ Equation (\ref{vlasovgamma}) together with 
Einstein's equations 
\begin{displaymath} 
G_{ab}:=R_{ab}-\frac{1}{2}Rg_{ab}=8\pi T_{ab}+\Lambda g_{ab}, 
\end{displaymath} 
then form the Einstein-Vlasov system. 
Here $G_{ab}$ is the Einstein tensor, $R_{ab}$ the Ricci tensor, $R$
is the scalar curvature and $\Lambda$ is the cosmological constant. In
most of this review we will assume that $\Lambda=0$ but section 2.3 is 
devoted to the case of non-vanishing cosmological constant (where also
the case of adding a scalar field is discussed). 
We now define the particle current density 
\begin{displaymath} 
N^a=-\int_{\mathbb{R}^{3}}f\,p^a|g|^{1/2}\frac{dp^1dp^2dp^3}{p_0}. 
\end{displaymath} 
Using normal coordinates based at a given point and assuming that $f$
is compactly supported it is not hard to see
that $T_{ab}$ is divergence-free which is a necessary compatability
condition since $G_{ab}$ is divergence-free by the Bianchi identities. 
A computation in normal coordinates also shows that $N^a$ is
divergence-free, which expresses the fact that the number of particles
is conserved. The definitions of $T_{ab}$ and $N^a$ immediately give us
a number of inequalities. If $V^a$ is a future directed timelike or
null vector then we have $N_aV^a\leq 0$ with equality if and only if
$f=0$ at the given point. Hence $N^a$ is always future directed
timelike if there are particles at that point. 
Moreover, if $V^a$ and $W^a$ are future directed timelike vectors then
$T_{ab}V^aW^b\geq 0,$ which is the dominant energy condition. 
If $X^a$ is a spacelike vector then $T_{ab}X^aX^b\geq 0.$ This is 
called the non-negative pressure condition. These last two conditions
together with the Einstein equations imply that $R_{ab}V^aV^b\geq 0$
for any timelike vector $V^a,$ which is the strong energy condition. 
That the energy conditions hold for Vlasov matter is one reason that
the Vlasov equation defines a well-behaved matter model in general
relativity. Another reason is the well-posedness theorem by
Choquet-Bruhat for the Einstein-Vlasov system that we will state below. 
Before stating that theorem we will first 
discuss the initial conditions imposed. 

The data in the Cauchy problem for the Einstein-Vlasov system consist
of the induced Riemannian metric $g_{ij}$ on the initial hypersurface $S$, the
second fundamental form $k_{ij}$ of $S$ and matter data $f_0$. 
The relations between a given initial data set 
$(g_{ij},k_{ij})$ on a three-dimensional manifold $S$ and the metric $g_{ij}$ 
on the spacetime manifold is that there exists an embedding $\psi$ of $S$ 
into the spacetime such that the induced metric and second fundamental form of 
$\psi(S)$ coincide with the result of transporting $(g_{ij},k_{ij})$
with $\psi$. 
For the relation of the distribution functions $f$ and $f_0$ we have to 
note that $f$ is defined on the mass shell. The initial condition
imposed is that the restriction of $f$ to the part of the mass shell
over $\psi(S)$  
should be equal to $f_0\circ (\psi^{-1},d(\psi)^{-1})\circ\phi$ where $\phi$ 
sends each point of the mass shell over $\psi(S)$, to its orthogonal 
projection onto the tangent space to $\psi(S)$. An initial data 
set for the Einstein-Vlasov system must satisfy the constraint 
equations, which read 
\begin{eqnarray}
R-k_{ij}k^{ij}+(\mbox{tr}k)^2&=&16\pi\rho,\label{constr1}\\ 
\nabla_{i}k^{i}_{l}-\nabla_{l}(\mbox{tr}k)&=&8\pi j_l.\label{constr2} 
\end{eqnarray} 
Here $\rho=T_{ab}n^an^b$ and $j^a=-h^{ab}T_{bc}n^c$ where $n^a$ is the
future directed unit normal vector to the initial hypersurface and
$h^{ab}=g^{ab}+n^an^b$ is the orthogonal projection onto the tangent
space to the initial hypersurface. In terms of $f_0$ we can express
$\rho$ and $j^l$ by ($j^a$ satisfies $n_aj^a=0$ so it can naturally be
identified with a vector intrinsic to $S$): 
\begin{eqnarray*}
\rho&=&\int_{\mathbb{R}^{3}}f\,p^ap_a|^{(3)}g|^{1/2}\frac{dp^1dp^2dp^3}
{1+p_jp^j},\\ 
j_l&=&\int_{\mathbb{R}^{3}}f\,p_l|^{(3)}g|^{1/2}\, dp^1dp^2dp^3. 
\end{eqnarray*} 
Here $|^{(3)}g|$ is the determinant of the induced Riemannian metric
on $S.$ 
We can now state the local existence theorem by Choquet-Bruhat~\cite{Ct} 
for the Einstein-Vlasov system. 
\begin{theorem} 
Let $S$ be a 3-dimensional manifold, $g_{ij}$ a smooth Riemannian
metric on $S,$ $k_{ij}$ a smooth symmetric tensor on $S$ and $f_0$ a
smooth non-negative function of compact support on the tangent bundle
$TS$ of $S$. Suppose that these objects satisfy the constraint
equations (\ref{constr1}) and (\ref{constr2}). Then there exists a
smooth spacetime $(M,g_{ab}),$ a smooth distribution function $f$ on
the mass shell of this spacetime, and a smooth embedding $\psi$ of $S$
into $M$ which induces the given initial data on $S$ such that
$g_{ab}$ and $f$ satisfy the Einstein-Vlasov system and $\psi(S)$ is a
Cauchy surface. Moreover, given any other spacetime $(M',g'_{ab}),$
distribution function $f'$ and embedding $\psi'$ satisfying these
conditions, there exists a diffeomorphism $\chi$ from an open
neighbourhood of $\psi(S)$ in $M$ to an open neighbourhood of
$\psi'(S)$ in $M'$ which satisfies $\chi\circ\psi=\psi'$ and carries
$g_{ab}$ and $f$ to $g'_{ab}$ and $f'$ respectively.  
\end{theorem} 
In this context we also mention that local existence has been proved
for the Yang-Mills-Vlasov system in~\cite{CN} and that this problem
for the Einstein-Maxwell-Boltzmann system is treated
in~\cite{BC}. This result is however not complete, the nonnegativity
of $f$ is left unanswered. The paper also raises the question on what
hypothesis from a physical point of view are reasonable on the
scattering kernel in curved spacetimes, an issue which does not seem 
fully understood. 

A main theme in the following sections is to discuss special cases
for which the local existence theorem can be extended to a global
one. There are interesting situations when this can be achieved, and
such global existence theorems are not known for Einstein's equations
coupled to other forms of phenomenological matter models, {\it i.e.} fluid
models (see, however,~\cite{Cu3}). 
In this context it should be stressed that the results in the
previous sections show that the mathematical understanding of kinetic
equations on a flat background space is well-developed. On the other
hand the mathematical understanding of fluid
equations on a flat background space (also in the absence of a
Newtonian gravitational field) is not satisfying. It 
would be desirable to have a better mathematical understanding of
these equations in the absence of gravity before coupling them to
Einstein's equations. This suggests that the Vlasov equation is
natural as matter model in mathematical general relativity. 
\section{Global existence theorems for the Einstein-Vlasov system} 
In general relativity two classes of initial data are distinguished. 
If an isolated matter distribution is studied, the data are called
asymptotically 
flat. The initial hypersurface is topologically $\mathbb{R}^{3}$ and (since far
away from the body one expects spacetime to be approximately 
flat) appropriate fall off conditions are
imposed. Roughly, a 
smooth data set $(g_{ij},k_{ij},f_0)$ on $\mathbb{R}^{3}$ is said to be
asymptotically flat 
if there exist global coordinates $x^i$ such that as $|x|$ tends to
infinity the components $g_{ij}$ in these coordinates tend to
$\delta_{ij},$ the components $k_{ij}$ tend to zero, $f_0$ has
compact support and certain weighted Sobolev norms of
$g_{ij}-\delta_{ij}$ and $k_{ij}$ are finite (see~\cite{Rl6}). 
The symmetry classes that admit asymptotically flatness are few. 
The important ones are spherically symmetric and axially symmetric 
spacetimes. One can also consider a case in which spacetime is
asymptotically flat except in one direction, namely cylindrical
spacetimes. A lot of work has been done on the spherically
symmetric case and this will be reviewed below. In the case of
cylindrical symmetry it has been shown that if singularities form then
the first one must occur at the axis of symmetry~\cite{F}. 

Spacetimes that possess a compact Cauchy hypersurface are called
cosmological spacetimes and data are accordingly given on a compact
3-manifold. In this case the whole universe is modelled and not only
an isolated body. In contrast to the asymptotically flat case,
cosmological spacetimes admit a large number of symmetry classes. This 
gives one the possibility to study interesting special cases for which the 
difficulties of the full Einstein equations are strongly reduced. 
We will discuss below cases for which the spacetime is characterized by 
the dimension of its isometry group together with the dimension of 
the orbit of the isometry group. 

\subsection{Spherically symmetric spacetimes} 
The study of the global properties of solutions to the spherically
symmetric Einstein-Vlasov system was initiated by Rein and Rendall in
1990. 
They chose to work in coordinates where the metric takes the form 
\begin{displaymath}
ds^{2}=-e^{2\mu(t,r)}dt^{2}+e^{2\lambda(t,r)}dr^{2}
+r^{2}(d\theta^{2}+\sin^{2}{\theta}d\varphi^{2}), 
\end{displaymath} 
where $t\in\mathbb{R},\, r\geq 0,\,\theta\in [0,\pi],\,\varphi\in
[0,2\pi].$ These are called Schwarzschild coordinates. 
Asymptotic flatness is expressed by the boundary conditions 
\begin{displaymath} 
\lim_{r\rightarrow\infty}\lambda(t,r)=\lim_{r\rightarrow\infty}\mu(t,r)
=0,\;\forall t\geq 0. 
\end{displaymath} 
A regular centre is also required and is guaranteed by the boundary
condition 
\begin{displaymath} 
\lambda(t,0)=0. 
\end{displaymath} 
With $$x=(r\sin\phi\cos\theta,r\sin\phi\sin\theta,r\cos\phi)$$ as 
spatial coordinate and $$v^j=p^j+(e^\lambda-1)\frac{x\cdot
  p}{r}\frac{x^j}{r}$$ as momentum coordinates the Einstein-Vlasov
system reads 
\begin{equation} 
\partial_{t}f+e^{\mu-\lambda}\frac{v}{\sqrt{1+|v|^{2}}}\cdot\nabla_{x}f
-(\lambda_{t}\frac{x\cdot v}{r}+e^{\mu-\lambda}\mu_{r}\sqrt{1+|v|^{2}})
\frac{x}{r}\cdot\nabla_{v}f=0,\label{Vlas} 
\end{equation} 
\begin{eqnarray} 
&\displaystyle e^{-2\lambda}(2r\lambda_{r}-1)+1=8\pi
r^2\rho,&\label{ee1}\\ 
&\displaystyle e^{-2\lambda}(2r\mu_{r}+1)-1=8\pi r^2 p.&\label{ee2} 
\end{eqnarray} 
The matter quantities are defined by 
\begin{eqnarray} 
\rho(t,x)&=&
\int_{\mathbb{R}^{3}}\sqrt{1+|v|^2}f(t,x,v)\;dv,\label{rho}\\ 
p(t,x)&=&\int_{\mathbb{R}^{3}}\left(\frac{x\cdot
    v}{r}\right)^{2}f(t,x,v)\;\frac{dv}{\sqrt{1+|v|^2}}.\label{p} 
\end{eqnarray} 
Let us point out that this system is not the full Einstein-Vlasov
system. The remaining field equations, however, can be derived from
these equations. See~\cite{RR1} for more details. 
Let the square of the angular momentum be denoted by $L,$ i.e. 
$$L:=|x|^2|v|^2-(x\cdot v)^2.$$ 
A consequence of spherical symmetry is that angular momentum 
is conserved along the characteristics of (\ref{Vlas}). 
Introducing the variable 
$$w=\frac{x\cdot v}{r},$$ the Vlasov equation for $f=f(t,r,w,L)$ 
becomes 
\begin{equation} 
\partial_{t}f+e^{\mu-\lambda}\frac{w}{E}\partial_{r}f
-(\lambda_{t}w+e^{\mu-\lambda}\mu_{r}E-
e^{\mu-\lambda}\frac{L}{r^3E})\partial_{w}f=0,\label{Vlas2} 
\end{equation} 
where 
$$E=E(r,w,L)=\sqrt{1+w^{2}+L/r^{2}}.$$ 
The matter terms take the form 
\begin{eqnarray} 
\rho(t,r)&=&\frac{\pi}{r^{2}}
\int_{-\infty}^{\infty}\int_{0}^{\infty}Ef(t,r,w,L)\;dwdL,\label{rho2}\\ 
p(t,r)&=&\frac{\pi}{r^{2}}\int_{-\infty}^{\infty}\int_{0}^{\infty}
\frac{w^{2}}{E}f(t,r,w,L)\;d
wdL.\label{p2} 
\end{eqnarray} 
Let us write down a couple of known 
facts about the system
(\ref{ee1}),(\ref{ee2}),(\ref{Vlas2}),(\ref{rho2}), and (\ref{p2}). 
A solution to the Vlasov 
equation can be written as:
\begin{equation}
f(t,r,w,L)=f_{0}(R(0,t,r,w,L),W(0,t,r,w,L),L), 
\label{repre} 
\end{equation} 
where $R$ and $W$ are solutions to the characteristic system,
\begin{eqnarray}
\frac{dR}{ds}&=&e^{(\mu-\lambda)(s,R)}\frac{W}{E(R,W,L)},\label{char1}\\ 
\frac{dW}{ds}&=&-\lambda_{t}(s,R)W-e^{(\mu-\lambda)(s,R)}\mu_{r}(s,R)E(R,W,L)
\label{char2}\\ 
&\phantom{hej}&+e^{(\mu-\lambda)(s,R)}\frac{L}{R^3E(R,W,L)},\nonumber 
\end{eqnarray} 
such that the trajectory $(R(s,t,r,w,L),W(s,t,r,w,L),L)$ goes 
through the point $(r,w,L)$ when $s=t$. 
This representation shows that $f$ is nonnegative for all $t\geq 0$ and that 
$f\leq\|f_0\|_{\infty}.$ 
There are two known conservation laws for the Einstein-Vlasov
system: conservation of the number of particles, 
\begin{displaymath} 
4\pi^{2}\int_{0}^{\infty}e^{\lambda(t,r)}
\left(\int_{-\infty}^{\infty}\int_{0}^{\infty}f(t,r,w,L)dLdw\right) dr, 
\end{displaymath} 
and conservation of the ADM mass 
\begin{equation} 
M:=4\pi\int_{0}^{\infty}r^{2}\rho(t,r)dr.\label{adm} 
\end{equation} 

Let us now review the global results concerning the Cauchy problem
that have been proved for 
the spherically symmetric Einstein-Vlasov system. As initial data we
take a spherically symmetric, nonnegative, and 
continuously differentiable function $f_0$ with compact support that
satisfies 
\begin{equation} 
4\pi^{2}\int_{0}^{r}\int_{-\infty}^{\infty}\int_{0}^{\infty}
Ef_{0}(r,w,L)dwdLdr<\frac{r}{2}.\label{ts} 
\end{equation} 
This condition guarantees that no trapped surfaces are present initially. 
In~\cite{RR1} it is shown that for such an initial datum there exists a
unique, continuously differentiable solution $f$ with $f(0)=f_0$ on
some right maximal interval $[0,T).$ If the solution blows up in
finite time, i.e. if $T<\infty$ then $\rho(t)$ becomes unbounded as
$t\rightarrow T.$ Moreover, a continuation criterion is shown that
says that a local solution can be extended to a global one provided 
the $v$-support of $f$ can be bounded on $[0,T)$ (in~\cite{RR1} they 
chose to work in the momentum variable $v$ rather than $w,L.$). 
This is analogous to the situation for the Vlasov-Maxwell system where
the function $Q(t)$ was introduced for the $v$-support. A control of
the $v$-support immediately implies that $\rho$ and $p$ are bounded in
view of (\ref{rho}) and (\ref{p}). In the Vlasov-Maxwell case the
field equations have a regularizing effect in the sense that 
derivatives can be expressed through spatial integrals, and it follows 
[GSt1] that the derivatives of $f$ also can be bounded if the 
$v$-support is bounded. For 
the Einstein-Vlasov system such a regularization is less clear, since 
e.g. $\mu_r$ depends on $\rho$ in a pointwise manner. However, certain 
combinations of second and first order derivatives of the metric
components can be expressed in
terms of matter components only--without derivatives (a
consequence of the geodesic deviation equation). This 
fact turns out to be sufficient for obtaining bounds
also on the derivatives of $f$ (see~\cite{RR1} for details). 
By considering initial data sufficiently close to zero, Rein 
and Rendall show that the $v$-support is bounded on 
$[0,T)$ and the continuation criterion then implies that $T=\infty.$ It
should be stressed that even for small data no global 
existence result like this one is known for any other phenomenological
matter model coupled to Einstein's equations. 
The resulting spacetime in~\cite{RR1} is geodesically 
complete and the components of the energy momentum tensor
as well as the metric quantities decay with certain algebraic rates in
$t.$ The mathematical method used by Rein and Rendall is inspired 
by the analogous small data result for the Vlasov-Poisson equation by 
Bardos and Degond~\cite{BD}. This should not be too surprising since for
small data the gravitational fields are expected to be small and a
Newtonian spacetime should be a fair approximation. In this context we
point out that in~\cite{RR2} it is proved that the Vlasov-Poisson
system is indeed the nonrelativistic limit of the spherically
symmetric Einstein-Vlasov system,
i.e. the limit when the speed of light $c\rightarrow \infty.$ 
(In~\cite{Rl3} this issue is studied in the asymptotically flat case
without symmetry assumptions.) 
Finally, we mention that there is an analogous small data result using
a maximal time coordinate~\cite{Rl6} instead of a Schwarzschild time
coordinate. In these coordinates trapped surfaces are allowed in
contrast to the Schwarzschild coordinates. 

The case with general data is more subtle. 
Rendall has shown~\cite{Rl7} that there exists data leading to singular
spacetimes as a consequence of Penrose's singularity theorem. 
This raises the question of what we mean by global existence for such 
data. The Schwarzschild time coordinate is expected to avoid the
singularity and by global existence we mean that solutions remain
smooth as Schwarzschild time tends to infinity. Even though spacetime 
might be only partially covered in Schwarzschild coordinates, a global
existence theorem for general data would nevertheless 
be very important since it is likely that it would constitute a
central step for proving weak cosmic censorship. Indeed, if this
coordinate system can be shown to cover the domain of outer
communications and if null infinity could be shown to be complete then 
weak cosmic censorship would follow. 
A partial investigation for general data in Schwarzschild coordinates
was done in~\cite{RRS1} and in maximal-isotropic coordinates
in~\cite{Rl6}. In Schwarzschild coordinates it is shown that 
if singularities form in finite time the first one must
be at the centre. More precisely, if $f(t,r,w,L)=0$ when $r>\epsilon$
for some 
$\epsilon>0$, and for all $t,w$ and $L,$ then the solution remains
smooth for all
time. This rules out singularities of the shell-crossing type, which can
be an annoying problem for other matter models ({\it i.e.} dust). The main
observation 
in~\cite{RRS1} is a cancellation property in the term $$\lambda_t w+
e^{\mu-\lambda}\mu_rE$$ in the characteristic equation (\ref{char2}). 
In~\cite{Rl6} a similar result is shown but here also an assumption
that one of the metric functions is bounded at the centre is
assumed. However, with this assumption the result follows in a more
direct way and the analysis of the Vlasov equation is not necessary 
which indicates that such a result might be true more generally. 
Recently, Dafermos and Rendall~\cite{DR} have shown a similar result
for the Einstein-Vlasov system in double-null coordinates. The main 
motivation for studying the system in these coordinates has its origin
from the method of proof of the cosmic censorship
conjecture for the Einstein-scalar field system by
Christodoulou~\cite{Cu2}. An essential part of his method is based on
the understanding of the formation of trapped surfaces
(~\cite{Cu5}). The presence of trapped surfaces (for the relevant
initial data) is then crucial in proving the completeness of future
null infinity in~\cite{Cu2}. In~\cite{D} the relation between trapped
surfaces and the completeness of null infinity was strengthened; a
single trapped surface or marginally trapped surface in the maximal
development implies completeness of null infinty. The theorem holds
true for any spherically symmetric matter spacetime if the matter
model is such that ``first'' singularities necessarily emanate from
the center (where the notion of ``first'' is tied to the casual
structure). The results in~\cite{RRS1} and in~\cite{Rl6} are not 
sufficient for concluding that the hypothesis of the matter needed in
the theorem in~\cite{D} is satisfied since they concern a portion of
the maximal development covered by particular coordinates. Therefore,
Dafermos and Rendall~\cite{DR} choose double-null coordinates which
cover the maximal development and they show that the mentioned 
hypothesis is satisfied for Vlasov matter. 

In~\cite{RRS2} a numerical study was undertaken of the Einstein-Vlasov
system in Schwarzschild coordinates. 
A numerical scheme originally used for the Vlasov-Poisson system 
was modified to the spherically symmetric Einstein-Vlasov system. 
It has been shown by Rein and Rodewis~\cite{RRo} that the numerical scheme has 
the desirable convergence properties. (In the Vlasov-Poisson case
convergence was proved in~\cite{Sc3}. See also~\cite{Ga}). 
The numerical experiments support the conjecture 
that solutions are 
singularity-free. This can be seen as evidence that weak cosmic
censorship holds for collisionless matter. It may even hold in
a stronger sense than in the case of a massless scalar field
(see~\cite{Cu1,Cu2}). There may be no naked singularities formed for
any regular 
initial data rather than just for generic data. This speculation is
based on the fact that 
the naked singularities that occur in scalar field collapse appear to
be associated with the existence of type II critical collapse, while
Vlasov matter is of type I. This is indeed the primary goal of 
their numerical investigation--to analyze critical collapse and
decide whether Vlasov matter is type I or type II. 

These different types of matter are defined as follows. Given small initial
data no black holes are expected to form and matter will disperse
(which has been proved for a scalar field~\cite{Cu4} and for Vlasov
matter~\cite{RR1}). 
For large data, black holes will form and consequently there
is a transition regime separating 
dispersion of matter and formation of black holes. If we introduce a 
parameter $A$ on the initial data such that for small $A$ dispersion
occurs and for large $A$ a black hole is formed, we get a critical value
$A_c$ separating these regions. If we take $A>A_c$ and denote by
$m_B(A)$ the mass of the black hole, then if 
$m_B(A)\rightarrow 0$ as $A\rightarrow A_c,$ we have type II matter,
whereas for type I matter 
this limit is positive and there is a mass gap. For more 
information on critical collapse we refer to the review paper by 
Gundlach~\cite{Gu}. 

For Vlasov matter there is an independent numerical
simulation by Olabarrieta and Choptuik~\cite{OC} (using a maximal time
coordinate) and their conclusion agrees with the one in~\cite{RRS2}. 
Critical collapse is related to self 
similar solutions; Martin-Garcia and Gundlach~\cite{MG} have presented a
construction of such solutions for the massless Einstein-Vlasov system 
by using a method based partially on numerics. Since such solutions
often are related to naked singularities, it is important to note that
their result is for the massless case (in which case there is no known
analogous result to the small data theorem in~\cite{RR1}) and their initial
data are not in the class that we have described above. 

We end this section with a discussion of the spherically symmetric
Einstein-Vlasov-Maxwell system, i.e. the case considered above with
charged particles. Whereas the constraint equations in the uncharged
case do not involve any problems to solve once the distribution
function is given (and satisfies (\ref{ts})) the charged case is more
challenging. In~\cite{NNR} it is shown that solutions to the
constraint equations do exist for the Einstein-Vlasov-Maxwell
system. In~\cite{NN} local existence is then shown together with a
continuation criterion and in~\cite{Nou} the regularity theorem
in~\cite{RRS1} is generalized to this case. 

\subsection{Cosmological solutions} 
In cosmology the whole universe is modelled and the ``particles'' in
the kinetic description are galaxies or even clusters of galaxies. 
The main goal again is to determine the global properties of the solutions
to the Einstein-Vlasov system. In order to do so, a 
global time coordinate $t$ must be found (global existence) and 
the asymptotic behaviour of the solutions when $t$ tends to its
limiting values has to be analyzed. This might correspond to 
approaching a singularity ({\it e.g.} the big bang singularity) or to a phase
of unending expansion. Since the general case is beyond the range of
current mathematical techniques, all known results are for cases 
with symmetry (see however~\cite{Ae} where to some extent global properties 
are established in the case without symmetry). 

There are several existing results on global time coordinates for 
solutions of the Einstein-Vlasov system. In the spatially homogeneous 
case it is natural to choose a Gaussian time coordinate based on
a homogeneous hypersurface. The maximal range of a Gaussian time
coordinate in a solution of the Einstein-Vlasov 
system evolving from homogenous data on a compact manifold 
was determined in~\cite{Rl4}. The range is 
finite for models of Bianchi IX and Kantowski-Sachs types. It is finite in 
one time direction and infinite in the other for the other Bianchi types. 
The asymptotic behaviour of solutions in the spatially homogeneous
case has been analyzed in~\cite{RT} and~\cite{RU}. In~\cite{RT}, the case of massless
particles is considered, whereas the massive case is studied in~\cite{RU}. 
Both the nature of the initial singularity and the phase of unlimited
expansion are analyzed. The main concern is the behaviour of 
Bianchi models I, II, and III. The authors compare their solutions
with the solutions to the corresponding perfect fluid models. A
general conclusion is that the choice of
matter model is very important since for all symmetry classes studied 
there are differences between the collisionless model and a perfect fluid
model, both regarding the initial singularity and the expanding phase. 
The most striking example is for the Bianchi II models, where they find
persistent oscillatory behaviour near the singularity, which is quite
different from the known behaviour of Bianchi type II perfect fluid
models. 
In~\cite{RU} it also is shown that solutions for massive particles are
asymptotic to solutions with massless particles near the initial
singularity. For Bianchi I and II it  also is proved that solutions
with massive particles are asymptotic to dust solutions at late times.
It is conjectured that the same holds true also for Bianchi III. This
problem is then settled by Rendall in~\cite{Rl8}. 

All other results presently available on the subject concern spacetimes 
that admit a group of isometries acting on two-dimensional spacelike 
orbits, at least after passing to a covering manifold. 
The group may be 
two-dimensional (local $U(1)\times U(1)$ or $T^2$ symmetry) 
or three-dimensional 
(spherical, plane, or hyperbolic symmetry). In all these cases, the 
quotient of spacetime by the symmetry group has the structure of a 
two-dimensional Lorentzian manifold $Q$. The orbits of the group action 
(or appropriate quotients in the case of a local symmetry) are called 
surfaces of symmetry. Thus, there is a one-to-one correspondence between 
surfaces of symmetry and points of $Q$.
There is a major difference between the cases where the symmetry group
is two- or three-dimensional. In the three-dimensional case 
no gravitational waves are admitted, in contrast to the two-dimensional
case. In the former case, the field equations reduce to ODEs while in
the latter their evolution part consists of nonlinear wave equations. 
Three types of time coordinates that have been studied
in the inhomogeneous case are CMC, areal, and conformal 
coordinates. A CMC time coordinate, $t$, is one where each hypersurface 
of constant time has constant mean curvature (CMC) and on each hypersurface 
of this kind the value of $t$ is the mean curvature of that slice.
In the case of areal coordinates, the time coordinate is a function of
the area of the surfaces of symmetry ({\it e.g.} proportional to the area or 
proportional to the square root of the area). In the case of conformal  
coordinates, the metric on the quotient manifold $Q$ is conformally
flat. 

Let us first consider spacetimes $(M,g)$ admitting a three-dimensional
group of isometries. 
The topology of $M$ is assumed to be 
$\mathbb{R}\times S^1\times F,$ with $F$ a compact two-dimensional 
manifold. The universal covering $\hat{F}$ of $F$ induces a spacetime 
$(\hat{M},\hat{g})$ by $\hat{M}=\mathbb{R}\times S^1\times\hat{F}$ and
$\hat{g}=p^{*}g$, where $p:\hat{M}\rightarrow M$ is the canonical
projection. A three-dimensional group $G$ of isometries is assumed to
act on $(\hat{M},\hat{g}).$ If $F=S^2$ and $G=SO(3)$, then $(M,g)$ is
called spherically symmetric; if $F=T^2$ and $G=E_2$ (Euclidean group),
then $(M,g)$ is called plane symmetric; and if $F$ has genus greater 
than one and the connected component of the symmetry group $G$ of the
hyperbolic plane $H^2$ acts isometrically on $\hat{F}=H^2$, then
$(M,g)$ is said to have hyperbolic symmetry. 

In the case of spherical symmetry the existence of one compact CMC 
hypersurface implies that the whole spacetime can be covered by a 
CMC time coordinate that takes all real values~\cite{Rl1,BR}. 
The existence of one compact CMC hypersurface in this case
was proved later by Henkel~\cite{He1} using the concept of 
prescribed mean curvature (PMC) foliation. 
Accordingly this gives a complete picture in the spherically 
symmetric case regarding CMC foliations. 
In the case of areal coordinates, Rein~\cite{Rn2} has shown, under a size
restriction on the initial data, 
that the past of an initial hypersurface can be covered. In
the future direction it is shown that areal coordinates break down in
finite time. 

In the case of plane and hyperbolic symmetry, Rendall and 
Rein showed in~\cite{Rl1} and~\cite{Rn2}, respectively, 
that the existence results (for CMC time
and areal time) 
in the past direction for spherical symmetry also hold for 
these symmetry classes. The global CMC foliation results 
to the past imply that the past singularity is a crushing singularity since
the mean curvature blows up at the singularity. In addition, Rein also
proved in his special case with small initial data that 
the Kretschmann curvature scalar blows up when the singularity is
approached. 
Hence, the singularity is both a crushing and a curvature singularity
in this case. 
In both of these works the question of global existence to 
the future was left open.
This gap was closed in~\cite{ARR}, and global 
existence to the future was established in both CMC and areal 
coordinates. The global existence result for CMC time is partly a 
consequence of the global existence theorem in areal coordinates,
together with a theorem by Henkel~\cite{He1} that shows that there exists 
at least one hypersurface with (negative) constant mean curvature. 
Also, the past direction was analyzed in areal coordinates
and global existence was shown without any smallness condition
on the data. It is, however, not concluded if the past 
singularity in this more 
general case without the smallness condition on the data is a curvature
singularity as well. 
The question whether the areal time coordinate, which is positive by 
definition, takes all values in the range $(0,\infty)$ or only in
$(R_0,\infty)$ for some positive $R_0$ is also left open. 
In the special case in~\cite{Rn2}, it is indeed
shown that $R_0=0,$ but there is an example 
for vacuum spacetimes in the more general case of 
$U(1)\times U(1)$ symmetry 
where $R_0>0.$ This question was resolved by Weaver~\cite{Wea}. She 
proves that if spacetime contains Vlasov matter (i.e. $f\not= 0$) then
$R=0.$ Her result applies to a more general case which we now turn to. 

For spacetimes admitting a two-dimensional isometry group the first
study was done 
by Rendall~\cite{Rl2} in the case of local $U(1)\times U(1)$ symmetry (or
local $T^2$ symmetry). For a discussion of the topologies of 
these spacetimes we refer to the original paper. 
In the model case the spacetime is topologically of the form 
$\mathbb{R}\times T^3,$ and to simplify our discussion later on we
write down the metric in areal coordinates for this type of spacetime: 
\begin{eqnarray} 
&g=\mbox{e}^{2(\eta-U)}(-\alpha dt^{2}+d\theta^{2}) 
+\mbox{e}^{-2U}t^{2}[dy+Hd\theta+Mdt]^{2}&\nonumber\\ 
&+\mbox{e}^{2U}[dx+Ady+(G+AH)d\theta+(L+AM)dt]^{2}.&\label{areal} 
\end{eqnarray} 
Here the metric coefficients $\eta,U,\alpha,A,H,L$ and $M$ depend on
$t$ and $\theta$ and $\theta,x,y\in S^1.$ In~\cite{Rl2} CMC coordinates 
are considered rather than areal coordinates. The CMC- and the areal
coordinate foliations are both geometrically based time foliations. 
The advantage with a CMC 
approach is that the definition of a CMC hypersurface does not depend
on any symmetry assumptions and it is possible that CMC foliations
will exist for rather general spacetimes. 
The areal coordinate foliation, on the other hand, is 
adapted to the symmetry of spacetime but it has analytical advantages
that we will see below. 

Under the hypothesis that there exists at least one CMC hypersurface,
Rendall proves, without any smallness condition on the data, 
that the past of the given CMC hypersurface 
can be globally foliated by CMC hypersurfaces and that the mean
curvature of these hypersurfaces blows up at the past
singularity. Again, the future direction was left open. The result
in~\cite{Rl2} holds for Vlasov matter and for matter described by a
wave map 
(which is not a phenomenological matter model). That the choice of
matter model is important was shown by Rendall~\cite{Rl5} who gives a
non-global 
existence result for dust, which leads to examples of spacetimes~\cite{IsR}
that are not covered by a CMC foliation. 

There are several possible subcases to the $U(1)\times U(1)$ symmetry class. 
The plane case where
the symmetry group is three-dimensional is one subcase and the form of 
the metric in areal coordinates is obtained by letting $A=G=H=L=M=0$
and $U=\log{t}/2$ in (\ref{areal}). Another subcase, which still 
admits only two Killing fields (and which includes plane symmetry as a
special case), is Gowdy symmetry. It is obtained by letting $G=H=L=M=0$
in (\ref{areal}). In~\cite{An4}, the author considers Gowdy symmetric
spacetimes with Vlasov matter. It is proved that the entire   
maximal globally hyperbolic spacetime can be foliated by constant
areal time slices for arbitrary (in size) initial data. The areal 
coordinates are used in a direct way for showing global existence to
the future whereas the analysis for the past direction is carried out
in conformal coordinates. These coordinates are not fixed to the
geometry of spacetime and it is not clear that the entire past has 
been covered. A chain of geometrical arguments then shows that 
areal coordinates indeed cover the entire spacetime. This method 
was applied to the problem on hyperbolic and plane symmetry in~\cite{ARR}. 
The method in~\cite{An4} was in turn inspired by the work~\cite{BCIM}
for vacuum 
spacetimes where the idea of using conformal coordinates in the past 
direction was introduced. As pointed out in~\cite{ARR}, the result by
Henkel~\cite{He2} guarantees the existence of one CMC hypersurface in
the Gowdy case and, together with the global areal foliation
in~\cite{An4}, it follows that 
Gowdy spacetimes with Vlasov matter can be globally covered by CMC
hypersurfaces as well (also to the future). The general case of
$U(1)\times U(1)$ symmetry was considered in~\cite{ARW} where it is
shown that there exist global CMC and areal time foliations which
complete the picture. In this result as well as in the preceeding subcases
mentioned above the question whether or not the areal time coordinate
takes values in $(0,\infty)$ or in $(R,\infty),\; R>0,$ was left open. This
issue was solved by Weaver~\cite{Wea} where she concludes that $R=0$ if 
the distribution function is not identically zero initially. 

A number of important questions remain open. To analyse the nature of 
the initial singularity, which at present is known only for small
initial data in the case considered in~\cite{Rn2}, would be very
interesting. The question of the asymptotics in the future direction
is also an important issue where very little is known. The only
situation where a result has been obtained is in the case with
hyperbolic symmetry. Under a certain size restriction on the initial
data, Rein~\cite{Rn5} shows future geodesic completeness. However, in
models with a positive cosmological constant more can be said. 

\subsection{Cosmological models with a cosmological constant or a scalar field} 
The present cosmological observations indicate that the expansion of
the universe is accelerating and this has influenced the theoretical
studies in the field during the last years. One way to produce models
with accelerated expansion is to choose a positive cosmological 
constant in the Einstein equations. Another way is to include a
nonlinear scalar field among the matter fields. In this section we
will review the recent results for the Einstein-Vlasov system where a 
cosmological constant or a linear- or nonlinear scalar field have been 
included into the model. 

As in the previous section we start with the models with highest 
degree of symmetry, i.e. the locally spatially homogeneous models. In 
the case of a positive cosmological constant Lee~\cite{Le1} has shown 
global existence as well as future causal geodesic completeness for
initial data which have Bianchi symmetry. She also obtains the time
decay of the components of the energy momentum tensor as $t\to\infty.$ 
The past direction for some spatially homogeneous models is considered
in~\cite{T}. Existence back to the initial singularity is proved and
the case with a negative cosmological constant is discussed. 
In~\cite{Le2} Lee considers the case with a nonlinear
scalar field 
coupled to Vlasov matter. The form of the energy momentum then reads 
\begin{equation}
T_{\alpha\beta}=T_{\alpha\beta}^{\mbox{Vlasov}}+
\nabla_{\alpha}\phi\nabla_{\beta}\phi-(\frac{1}{2}\nabla^{\gamma}\phi\nabla_{\gamma}\phi+V(\phi))g_{\alpha\beta}.\label{TVNS} 
\end{equation}
Here $\phi$ is the scalar field and $V$ is a potential and the Bianchi
identities lead to the following equation for the scalar field 
\begin{equation}
\nabla^{\gamma}\phi\nabla_{\gamma}\phi=V'(\phi).\label{pot}
\end{equation} 
Under the assumption that $V$ is nonnegative and $C^2$ global
existence to the future is obtained and if the potential is restricted
to the form $$V(\phi)=V_{0}e^{-c\Phi},$$ where $0<c<4\sqrt{\pi}$ then
future geodesic completeness is proved. 

In the previous section we discussed the situation when spacetime
admits a three dimensional group of isometries and we distinguished
three cases: plane, spherical and hyperbolic symmetry. 
In area time coordinates the metric takes the form 
$$ds^2=-e^{\mu(t,r)}dt^2+e^{\lambda(t,r)}dr^2+t^2(d\theta^2+\mbox{sin}_{\kappa}^{2}\theta d\varphi^{2}),$$ 
where $k=0,-1,+1$ correspond to the plane, spherical and hyperbolic
case respectively and where $\mbox{sin}_{0}\theta=1, 
\mbox{sin}_{1}\theta=\sin{\theta}\mbox{ and
}\mbox{sin}_{-1}\theta=\sinh{\theta}.$ 
In~\cite{TR} 
the Einstein-Vlasov system with a positive cosmological constant is
investigated in the future (expanding) direction in the case of plane
and hyperbolic symmetry. The authors prove global existence to
the future in these coordinates and they also show future geodesic
completeness. The positivity of the cosmological constant is crucial
for the latter result. Recall that in the case of $\Lambda=0$ future
geodesic completenss has only been established for hyperbolic symmetry
under a smallness condition of the initial data~\cite{Rn5}. Finally a
form of the cosmic no hair conjecture is obtained for 
this class of spacetimes. Indeed, they show that the de Sitter
solution acts as a model for the dynamics of the solutions by proving
that the generalized Kasner exponents tend to $1/3$ as $t\to\infty$
which in the plane case is the de Sitter solution. The remaining case
of spherical symmetry is analysed in~\cite{TN}. Recall that when
$\Lambda=0,$ Rein~\cite{Rn2} showed that solutions can only exist for
finite time in the future direction in area time coordinates. By
adding a positive cosmological constant global existence to the future
is shown to hold true if initial data is given on $t=t_0$ where
$t_{0}^2>1/\Lambda.$ The asymptotic behaviour of the matter terms is
also investigated and slightly stronger decay estimates are obtained
in this case compared to the case of plane and hyperbolic symmetry. 

The results discussed so far in this section have concerned the future 
time direction and a positive cosmological constant. The past
direction with a negative cosmological constant is analysed
in~\cite{T} where it is shown that for plane and spherical symmetry
the areal time coordinate takes all positive
values which is in analogy with Weaver's~\cite{Wea} result for $\Lambda=0.$
If initial data is restricted by a
smallness condition the theorem is proven also in the hyperbolic case
and for such data the result of the theorem holds true in all of the
three symmetry classes when the cosmological constant is
positive. The early-time asymptotics in the case of small
initial data is also analysed and is shown to be Kasner-like. 

In~\cite{TNR} the Einstein-Vlasov system with a linear scalar field is
analysed in the case of plane, spherical and hyperbolic symmetry. Here
the potential $V$ in (\ref{TVNS}) and (\ref{pot}) is zero. A local
existence theorem and a continuation criterion, involving bounds on
derivatives of the scalar field in addition to a bound on the support
of one of the moment variables, is proven. For the Einstein-scalar
field system, i.e. when $f=0,$ the continuation criterion is shown to
be satisfied in the future direction and global existence follows in
that case.

\section{Stationary solutions to the Einstein-Vlasov system} 
Equilibrium states in galactic dynamics can be described as stationary 
solutions of the Einstein-Vlasov system, or of the Vlasov-Poisson
system in the Newtonian case. Here, we will consider the former case
for which only static, spherically symmetric solutions have been
constructed, but we mention that in the latter case also, stationary
axially symmetric solutions have been found by Rein~\cite{Rn6}. 

In the static, spherically symmetric case, the problem can be formulated
as follows. 
Let the space-time metric have the form 
\begin{displaymath}
ds^{2}=-e^{2\mu(r)}dt^{2}+e^{2\lambda(r)}dr^{2}
+r^{2}(d\theta^{2}+\sin^{2}{\theta}d\varphi^{2}), 
\end{displaymath} 
where $r\geq 0,\,\theta\in [0,\pi],\,\varphi\in
[0,2\pi].$ As before, asymptotic flatness is expressed by the boundary
conditions  
\begin{displaymath} 
\lim_{r\rightarrow\infty}\lambda(r)=\lim_{r\rightarrow\infty}\mu(r)
=0,\;\forall t\geq 0,
\end{displaymath}
and a regular centre requires 
\begin{displaymath} 
\lambda(0)=0. 
\end{displaymath} 
Following the notation in section 2.1, the time-independent
Einstein-Vlasov system reads 
\begin{equation} 
e^{\mu-\lambda}\frac{v}{\sqrt{1+|v|^2}}\cdot\nabla_{x}f
-\sqrt{1+|v|^2}e^{\mu-\lambda}\mu_{r}
\frac{x}{r}\cdot\nabla_{v}f=0,\label{Vlas3} 
\end{equation} 
\begin{eqnarray} 
&\displaystyle e^{-2\lambda}(2r\lambda_{r}-1)+1=8\pi r^2\rho,&\label{ee12}\\ 
&\displaystyle e^{-2\lambda}(2r\mu_{r}+1)-1=8\pi r^2 p.&\label{ee22} 
\end{eqnarray} 
The matter quantities are defined as before: 
\begin{eqnarray} 
\rho(x)&=&
\int_{\mathbb{R}^{3}}\sqrt{1+|v|^{2}}f(x,v)\;dv,\label{rho3}\\ 
p(x)&=&\int_{\mathbb{R}^{3}}\left(\frac{x\cdot
    v}{r}\right)^{2}f(x,v)\;\frac{dv}{\sqrt{1+|v|^{2}}}.\label{p3} 
\end{eqnarray} 
The quantities 
\begin{displaymath} 
E:=e^\mu(r)\sqrt{1+|v|^{2}},\,\,L=|x|^2|v|^{2}-(x\cdot v)^2=|x\times v|^2, 
\end{displaymath} 
are conserved along characteristics. $E$ is the particle energy and $L$
is the angular momentum squared. If we let $$f(x,v)=\Phi(E,L),$$ for
some function $\Phi$ the Vlasov equation is automatically 
satisfied. The form of $\Phi$ is usually restricted to 
\begin{equation} 
\Phi(E,L)=\phi(E)(L-L_0)^l,\label{pol}
\end{equation} 
where $l>-1/2$ and $L_0\geq 0.$ If $\phi(E)=(E-E_0)^k_{+},\,k>-1,$ for
some 
positive constant $E_0$ this is called the polytropic 
ansatz. The case of isotropic pressure is obtained by letting $l=0$ so 
that $\Phi$ only depends on $E$. We refer to~\cite{Rn3} for information on 
the role of $L_0$. 

In passing, we mention that for the Vlasov-Poisson system it
has been shown~\cite{BFH} that every static spherically symmetric solution must
have the form $f=\Phi(E,L).$ This is referred to as Jeans' 
theorem. It was an open question for some time to decide whether or
not this was 
also true for the Einstein-Vlasov system.  This was settled in 1999 by 
Schaeffer~\cite{Sc4}, who found solutions that do not have this particular
form globally on phase space, and consequently, Jeans' theorem is not valid in the relativistic 
case. 
However, almost all results in this field rest on this ansatz. 
By inserting the ansatz for $f$ in the matter quantities 
$\rho$ and $p$, a nonlinear system for $\lambda$ and $\mu$ is
obtained, in which
\begin{eqnarray*}
&\displaystyle e^{-2\lambda}(2r\lambda_{r}-1)+1=8\pi r^2G_{\Phi}(r,\mu),&\\ 
&\displaystyle e^{-2\lambda}(2r\mu_{r}+1)-1=8\pi r^2 H_{\Phi}(r,\mu),& 
\end{eqnarray*} 
where 
\begin{eqnarray*} 
G_{\Phi}(r,\mu)&=&
\frac{2\pi}{r^2}\int_1^{\infty}\int_0^{r^2(\epsilon^2-1)}\Phi(e^{\mu(r)}\epsilon,L)\frac{\epsilon^2}{\sqrt{\epsilon^2-1-L/r^2}}\;dLd\epsilon,\\ 
H_{\Phi}(r,\mu)&=&\frac{2\pi}{r^2}\int_1^{\infty}\int_0^{r^2(\epsilon^2-1)}\Phi(e^{\mu(r)}\epsilon,L)\sqrt{\epsilon^2-1-L/r^2}\;dLd\epsilon. 
\end{eqnarray*} 

Existence of solutions to this system was first proved in the case 
of isotropic pressure in~\cite{RR3} and then extended to the general case
in~\cite{Rn3}. 
The main problem is then to show that the resulting solutions 
have finite (ADM) mass and compact support. This is accomplished in~\cite{RR3} for a polytropic ansatz with isotropic pressure and in~\cite{Rn3} for 
a polytropic ansatz with possible anisotropic pressure. They 
use a perturbation argument based on the fact that the Vlasov-Poisson
system is the limit of the Einstein-Vlasov system as the speed of
light tends to infinity~\cite{RR2}. Two types of solutions are constructed, 
those with a regular centre (~\cite{RR3} and~\cite{Rn3}), and those (~\cite{Rn3}) with a 
Schwarzschild singularity in the centre. 
In~\cite{RR4} Rendall and Rein go beyond the polytropic ansatz and assume
that $\Phi$ satisfies 
\begin{displaymath} 
\Phi(E,L)=c(E-E_0)^k_{+}L^l+O((E_0-E)_{+}^{\delta+k})L^l\mbox{ as
  }E\rightarrow E_0, 
\end{displaymath} 
where $k>-1,\,l>-1/2,\,k+l+1/2>0,\,k<l+3/2.$ They show that this
assumption is sufficient for obtaining steady states with finite mass
and compact support. The result is obtained in a more direct way and
is not based on the perturbation argument mentioned above. Their
method is inspired by a work on stellar models by Makino~\cite{M}, in which he
considers steady states of the Euler-Einstein system. 
In~\cite{RR4} there is also an interesting discussion about steady states
that appear in the astrophysics literature. They show that 
their result applies to most of these steady states,
which proves that they have
the desirable property of finite mass and compact support. 

All solutions described so far have the property that the support of
$\rho$ contains a ball about the centre. In~\cite{Rn4} Rein shows that
there exist steady states whose support is a finite, spherically
symmetric shell, so that they have a vacuum region in the centre. 

At present, there are almost no known results concerning the stability
properties of the steady states to the Einstein-Vlasov system. 
In the Vlasov-Poisson case, however, the nonlinear stability of
stationary solutions has been investigated by Guo and Rein~\cite{GR} using
the energy-Casimir method. In the Einstein-Vlasov case, Wolansky~\cite{Wo}
has applied the energy-Casimir method and obtained some insights but
the theory in this case is much less developed than in the
Vlasov-Poisson case and the stability problem is essentially open. 

\section{Acknowledgements} 
I would like to thank Alan Rendall for helpful suggestions.


\end{document}